\documentclass[final,5p,times,twocolumn]{elsarticle}
\usepackage{natbib}
\usepackage[colorlinks,allcolors=blue]{hyperref}

\usepackage{times}
\usepackage{graphics}
\usepackage{rotating}
\usepackage{verbatim}
\usepackage{xcolor}
\usepackage{booktabs}
\usepackage{threeparttable}
\usepackage{tablefootnote}
\usepackage{textcomp}
\usepackage{amsmath}

\begin{document}

\begin{frontmatter}



\title{A comprehensive rotational study of astronomical iso-pentane within 84 to 111 GHz}

\author[inst1]{Anshika Pandey*}

\ead{3anshika.1@gmail.com}
\affiliation[inst1]{organization={Department of Physics, Institute of Science, Banaras Hindu University, Varanasi-221005, India}}
\author[inst1]{Satyam Srivastav}
\author[inst1]{Akant Vats}
\author[inst1]{Amit Pathak and K. A. P. Singh}

\cortext[cor1]{Corresponding author}

\begin{abstract}
The rotational line survey by ALMA (Atacama Large Millimeter/submillimeter Array) recently revealed the presence of i-C$_{3}$H$_{7}$CN (i-PrCN) and n-C$_{3}$H$_{7}$CN (n-PrCN) in 3-mm atmospheric window between 84 to 111 GHz towards the hot core region Sagittarius B2(N) (Sgr B2(N)). This was the first interstellar detection of a linear straight chain molecule. In this light, we report the rotational spectra of C$_{5}$H$_{12}$ isomeric group in the same frequency range. We performed quantum chemical calculations for spectroscopic parameters. The pure rotational spectrum of the species has been simulated using the PGOPHER program. The rotational spectrum of this molecule makes it a good candidate for future astronomical detections since the radio lines can be calculated to very high accuracy in mm/sub-mm wave region.     

\end{abstract}




\end{frontmatter}


\section{Introduction}
Rotational spectroscopy is the most precise and dependable source for structural information of gas-phase molecules \citep{dome1992}, which has led to the discovery of more than 240 chemical species in the interstellar medium (ISM) \citep{tennyson2019,herbst2001}. The study of small to medium-sized molecules in the millimeter and sub-millimeter wavelength range has opened prospects for the detection of new chemical species \citep{koerber2009,thorwirth2008}. In general, the molecules discovered in space are examined first in the laboratory using rotational spectroscopy. Following the calculation of accurate rotational constants and centrifugal distortion parameters, line positions for a detectable molecule in the ISM can be predicted (see, for example, Refs. \cite{mcmahon2003,brunken2007}. Since laboratory experiments are limited, theoretical predictions can greatly aid in the detection. Moreover, theoretical values for relevant spectroscopic observables can be used to guide experimental investigations \citep{2011Muller,BUCHANAN2021,NAKAJIMA2015,10.1093/mnras/stac3157}. The accurate rotational constant predictions allow laboratory spectroscopists to search in a relatively narrow frequency range, greatly facilitating the detection of unknown molecules \citep{cris2010}.

Aliphatic molecules are those with carbon atoms organized in open, chain-like configurations. While aliphatics with three or fewer carbon atoms must always have a simple straight-chain structure, aliphatics with four or more carbon atoms can have branched carbon chains. Normal propyl cyanide (n-C$_{3}$H$_{7}$CN/n-PrCN), one of the largest unbranched carbon-chain molecules identified in the ISM, was discovered at the Galactic Center hot-core region Sgr B2(N) \citep{Bell09}. Iso-propyl cyanide (i-C$_{3}$H$_{7}$CN/i-PrCN), the branched carbon-chain isomer of n-PrCN, was found with Atacama Large Millimeter/submillimeter Array (ALMA) telescope toward the same source as part of Exploring Molecular Complexity with ALMA (EMoCA) 3-mm line scan of Sgr B2(N) \citep{Bell14}. This was the first time a branched carbon-chain molecule was discovered in the ISM. \citet{pagani17} have also detected n- and i-C$_{3}$H$_{7}$CN in Orion. To date, the two isomeric forms of propyl cyanide together represent the largest molecules ever observed in a star-forming region \citep{Bell09,Bell14}.
\begin{figure*}
   \vspace{-3cm} \includegraphics[width=0.9\textwidth]{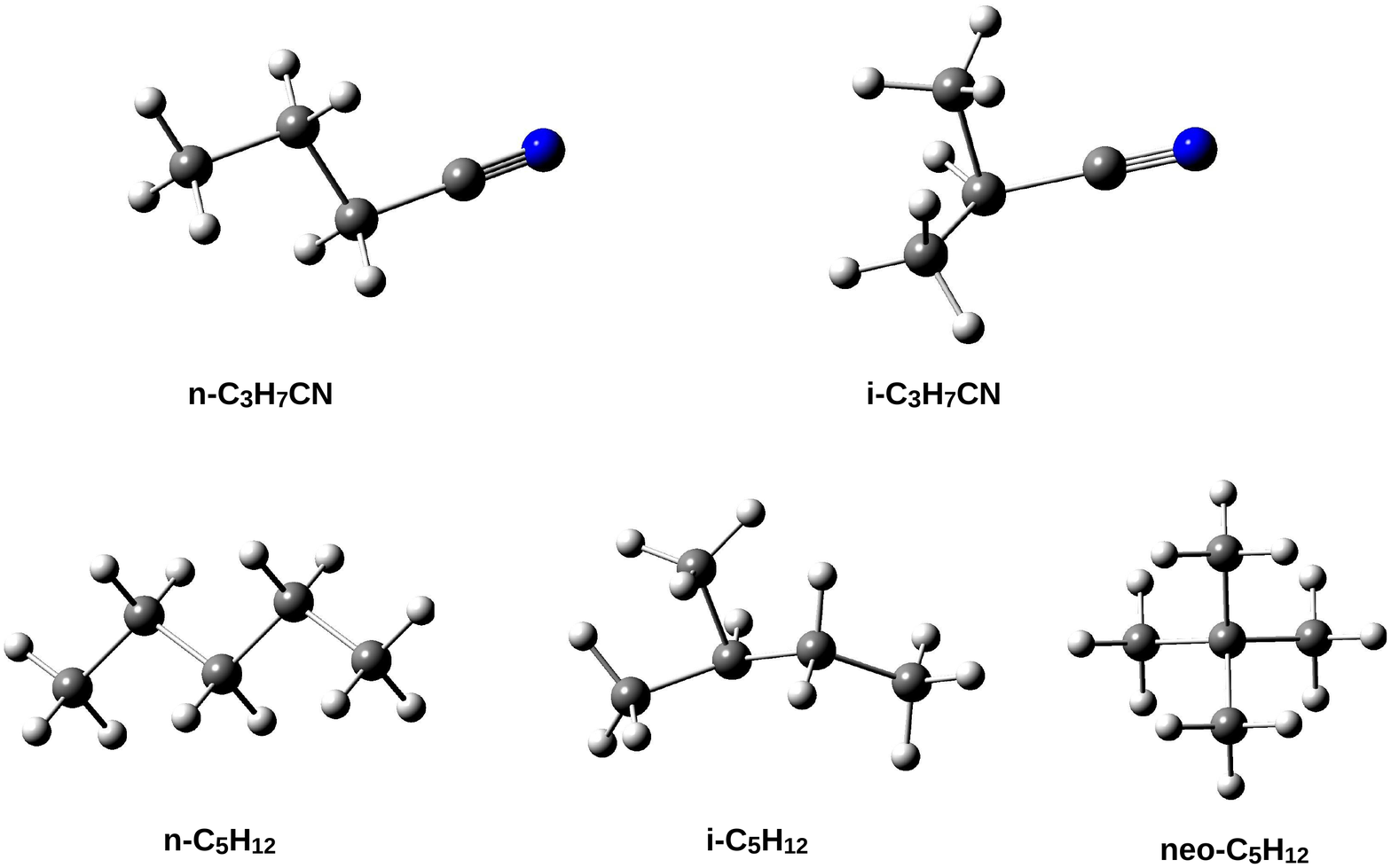}
    \caption{Structural representations of the straight-chain and branched forms of propyl cyanide (above) and pentane (below).}\label{Fig. 1}
\end{figure*}

Detection of iso-propyl cyanide in Sgr B2(N2) with a fractional abundance (of \begin{math}1.3\pm0.2\times10^{-8}\end{math}) in comparison to H$_{2}$, indicating an i-PrCN/n-PrCN ratio of \begin{math}0.40\pm0.06\end{math} \citep{Bell14}.
Hence, not only is the branched form of PrCN abundant, but it is also quite near to its straight-chain isomer, in terms of abundance. This implies that the chemical formation mechanisms of branched molecules may be competitive with those of their straight-chain counterparts, for even the simplest molecules of such structures.
The discovery of the two isomeric forms of propyl cyanide raises the question of what ratio of branched to straight-chain forms might be expected for other molecules of comparable size. To address this question, \citet{2017Garrod} expanded their earlier hot-core chemical kinetics model MAGICKAL, to incorporate a variety of reactions and processes related to the synthesis and destruction of butyl \& propyl cyanide, butane (C$_{4}$H$_{10}$), and pentane (C$_{5}$H$_{12}$). Butane comes in both normal and iso forms similar to propyl cyanide, while pentane can take three different forms; normal pentane iso-pentane and neo-pentane (n-, i- and neo-pentane). Fig. \ref{Fig. 1} shows structures of n-PrCN, i-PrCN and all forms of pentane.

The chemical model \citep{2017Garrod} takes into account mechanisms leading to the formation of butane at grain surfaces as well as cold-cloud gas-phase chemistry of straight-chain C$_{4}$ molecules. The model suggested the largest alkanes (butane and pentane) are produced largely by taking an H-atom out of a smaller alkane or by adding a H atom to an unsaturated hydrocarbon, followed by the addition of a methyl group. At all temperatures {(20, 50, 100, 200, 400 K)} and in each of the models (fast, medium, and slow warm up models), neo-pentane remains the least abundant and iso-pentane the most abundant form, though in different i/n ratios. Neo-pentane is less common than its structural isomers because its production requires the dissociation of i-C$_{4}$H$_{10}$ rather than the more prevalent smaller alkanes.
The higher abundance of iso-pentane, among the three forms of pentane, makes it more promising candidate for astronomical detection in the gas-phase. In the present work, we computed spectroscopic constants and simulated rotational spectrum of iso-pentane in mm/sub-mm wavelength region. We first established the accuracy of the employed theoretical methods on n-PrCN and i-PrCN before applying to the iso-pentane. We have also \textbf{\textcolor{red}}{listed} the pentane and its isomers' most stable conformers. 
\section{Computational Details}
Density Functional theory (DFT) is frequently used in the field of astrochemistry for studying small (up to 20 atoms) \citep{Bell09,Bell14,Muller2016,2022Sriv} as well as large (up to 400 atoms) molecules \citep{2007AcSpA..67..898P,2008ApJ...678..316B,2012ApJ...754...75R,2020Burago,2022Vats}. 
In earlier studies, DFT has been widely applied to estimate the molecular or spectroscopic constants for rotational spectroscopy \citep{NAKAJIMA2015,BUCHANAN2021,2011Muller,2012JPCA..116.5877B,10.1093/mnras/stac3157}. The commercially available program Gaussian 09 \citep{2013fris} red was used for the calculation of rotational parameters of n-PrCN, i-PrCN, and iso-pentane. The popular hybrid DFT methods B3LYP with basis sets 6-311++G(d,p)/aug-cc-pVTZ and MP2 with basis sets cc-pVDZ/aug-cc-pVDZ red were employed to optimize all the geometries followed by the anharmonic frequency calculations. The above-mentioned level of theories were used successfully in studying the pure rotational spectroscopy of interstellar molecules \citep{2011Muller,2016Das,2017Garrod}. After confirming the accuracy for n-PrCN and i-PrCN (test molecules) by matching the calculated data with the experiments, the level of theory with the best match was used for iso-pentane (target molecule).

The most stable conformer is thought to be the most likely candidate for astronomical detection. Before studying iso-pentane, we used a potential energy surface (PES) scan of dihedral angles to explore each possible conformer of the pentane's isomers. Since the MP2/cc-pVDZ level of theory is most suitable in conformational studies \citep{BUCHANAN2021,2011Muller}, the present work utilized the same level as well.
The PGOPHER suite of programmes \citep{2017Colin} was employed to simulate and analyze the rotational spectra of test and target molecules. The rotational-vibrational coupling, anharmonic and quartic centrifugal distortion corrections were considered for the rotational spectrum simulations at the second order vibrational perturbation theory (VPT2) level.  
\section{Results and Discussion}
\subsection{\textbf{Conformational study}}
The most stable conformer is anticipated to be the most likely candidate for astronomical discovery. In this approach, we first found the various conformers of the n/i/neo-pentane through a relaxed potential energy surface (PES) scan of dihedral angles. In previous astrochemical models, iso-pentane is found to be the most abundant isomer of pentane \citep{2017Garrod,2022Sriv}, however, for the sake of completeness, we discussed all the conformers of the isomers of pentane.
\begin{figure}
\centering
\vspace{-1.2cm}\includegraphics[width=0.5\textwidth]{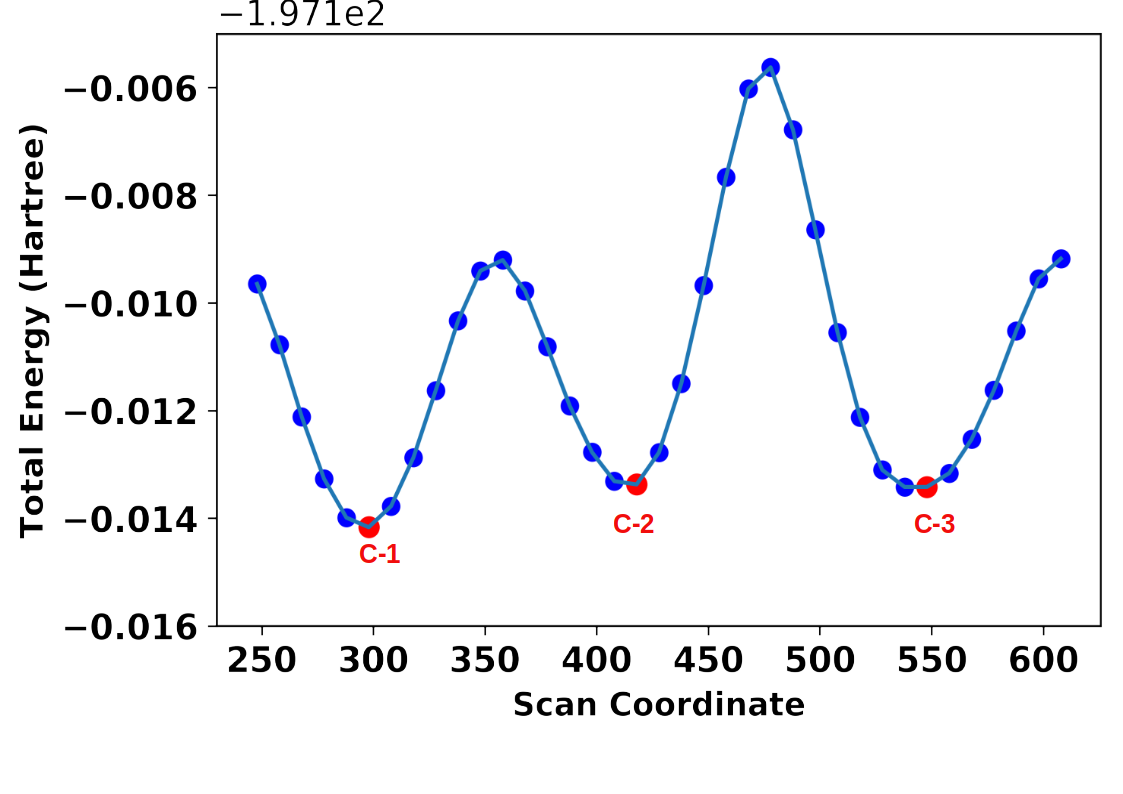}
\includegraphics[width=0.5\textwidth]{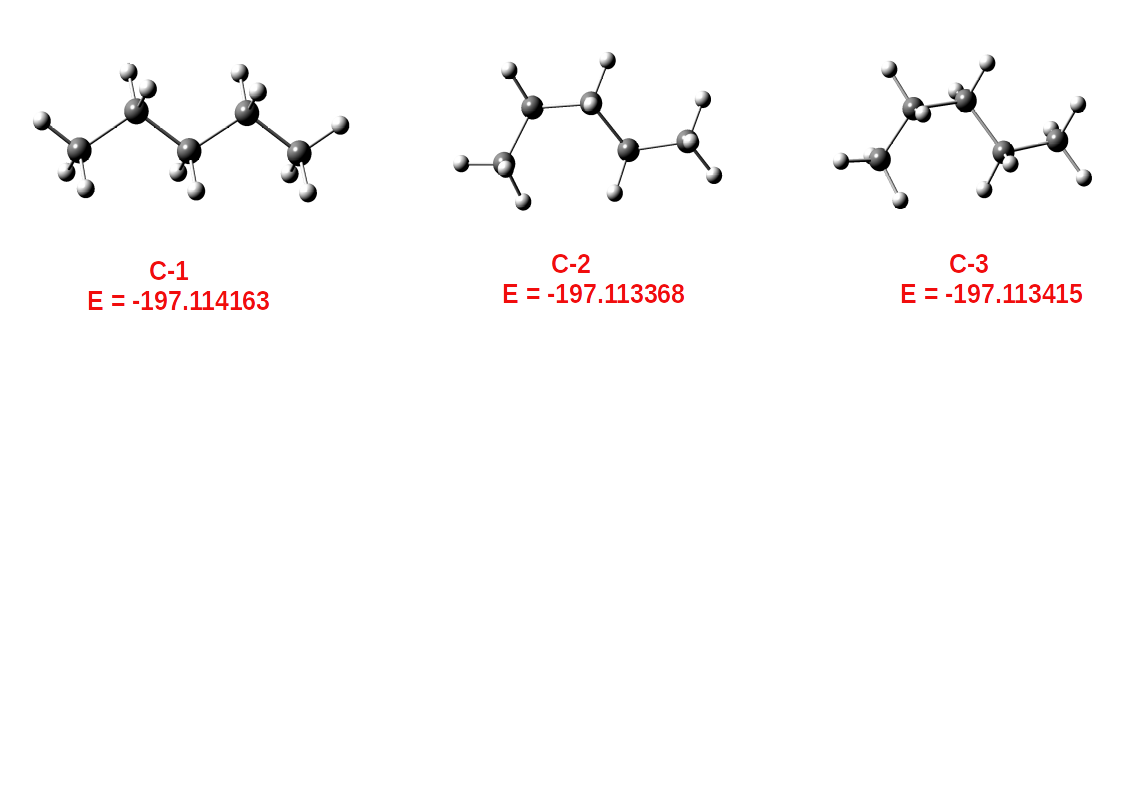}
\vspace{-3cm}\caption{Potential energy surface scan of the dihedral angle of n-pentane and corresponding ground state energy using the MP2/cc-pVDZ level of theory.}\label{fig:2}
 \end{figure}
\begin{figure}
\centering
\vspace{-1cm}\includegraphics[width=0.5\textwidth]{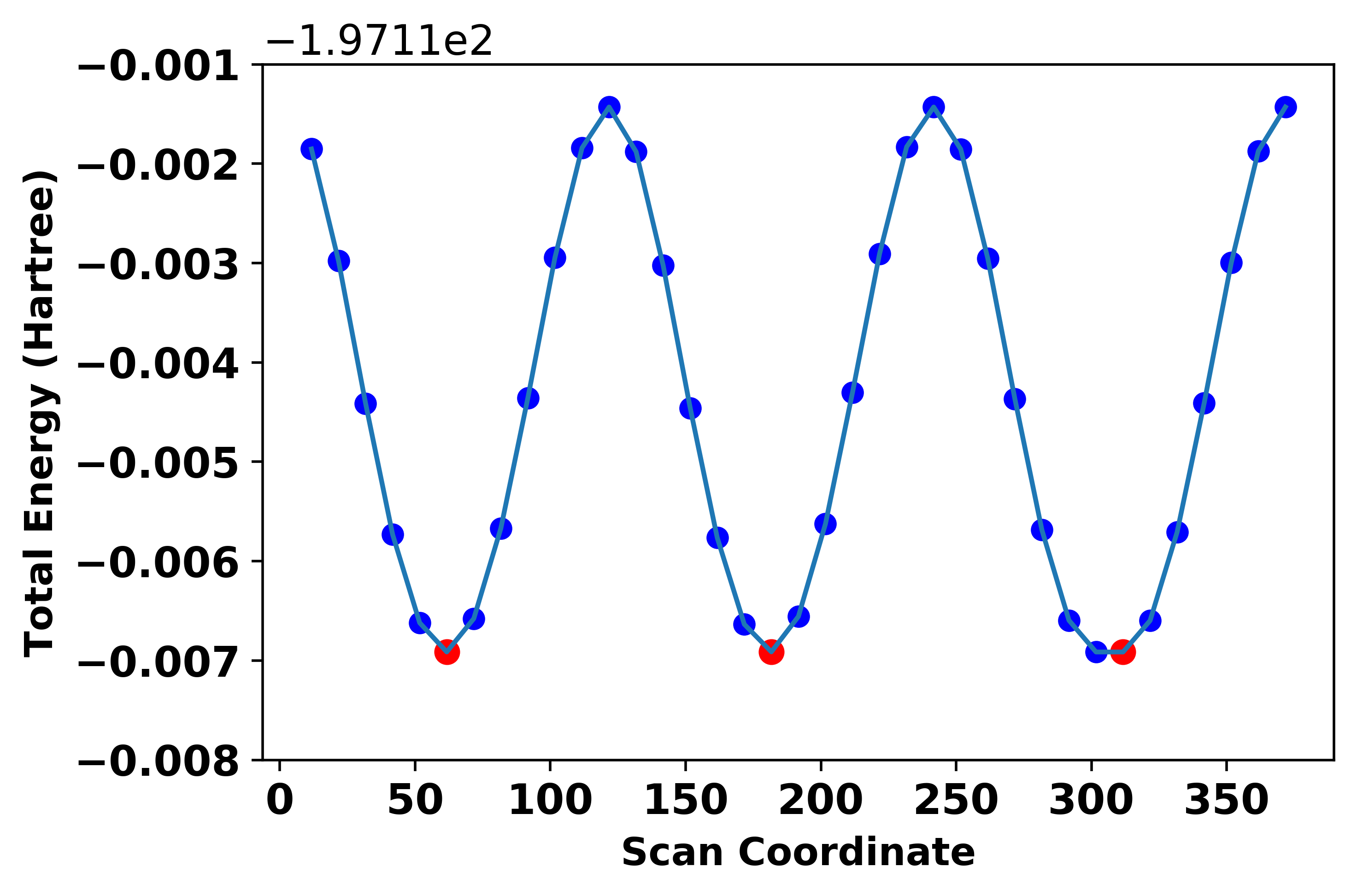}
\includegraphics[width=0.5\textwidth]{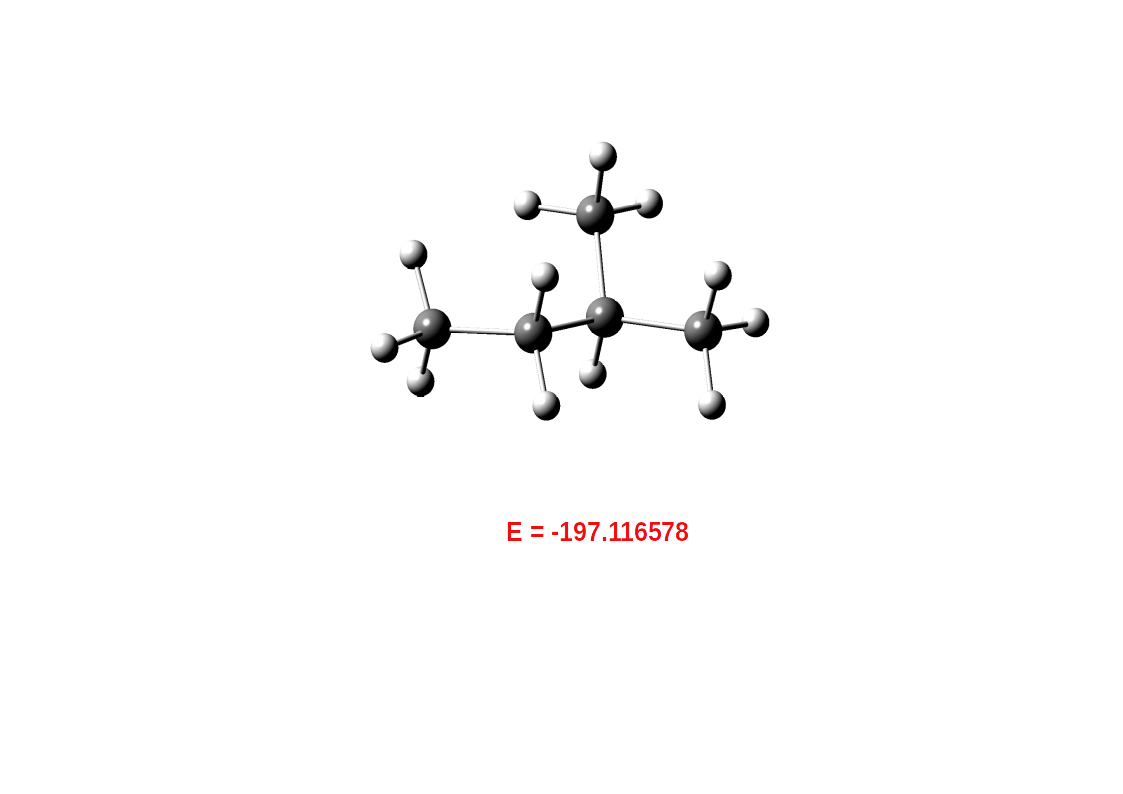}
\vspace{-2cm}\caption{Potential energy surface scan of the dihedral angle of iso-pentane with corresponding ground state energy using the MP2/cc-pVDZ level of theory}\label{fig:3}
\end{figure}
\begin{figure}
\centering
\includegraphics[width=0.5\textwidth]{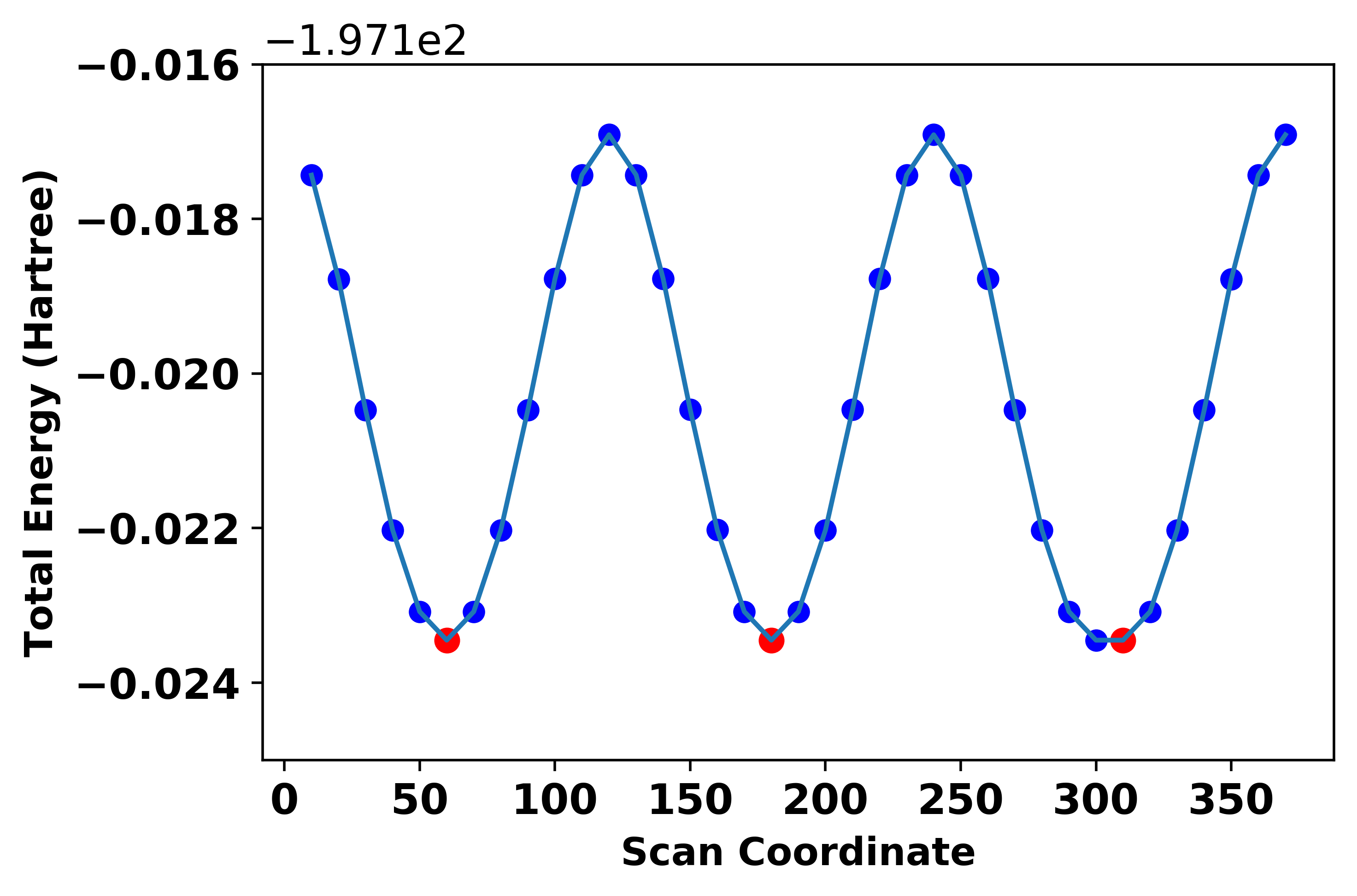}
\includegraphics[width=0.4\textwidth]{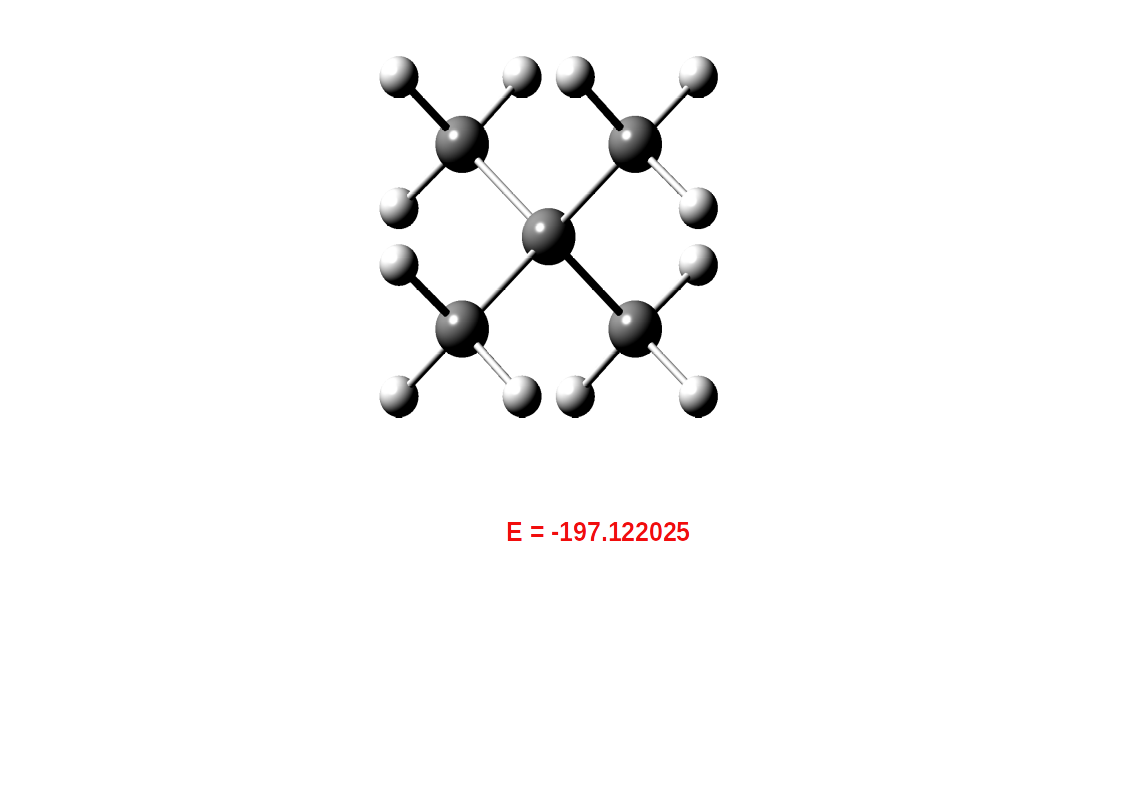}
\vspace{-1cm}\caption{Potential energy surface scan of the dihedral angle of neo-pentane with corresponding ground state energy using the MP2/cc-pVDZ level of theory}\label{fig:4}
\end{figure}

The PES scan results of the above mentioned species are shown in Figures 2, 3 \& 4. In Fig. 2, n-pentane having an internal rotation of
the CH$_{3}$ group, exists in the most stable structures at +60$^\circ$, -60$^\circ$ and 180$^\circ$ dihedral angles with the corresponding energies -197.1141, -197.1133 and -197.1134 a.u. respectively. 
\begin{table*}

\caption{Spectroscopic parameters in the ground state of n-propyl cyanide and i-propyl cyanide}
 \label{Tab 1}
 \resizebox{\textwidth}{!}{{\begin{tabular}{lccccccc}
    \hline
    \hline
 Parameters&\multicolumn{4}{c}{This work} & Experimental& Experimental\\ 
 \cline{2-5}
  &B3LYP&MP2&MP2&B3LYP&& uncertainty\\ 
  &6311++G(d,p)&cc-pVDZ&aug-cc-pVDZ&aug-cc-pVTZ &\\
 \hline
 \hspace{-1.8mm} n-PrCN &&&&\\
A$_{0}$$\rm$ (MHz)&23985.4623&23482.1698&23359.59157&23571.4914&23668.3193&14\\
B$_{0}$$\rm$ (MHz)&2249.5510&2235.6388&2237.4096&2267.7725&2268.1469&15\\
C$_{0}$$\rm$ (MHz)&2138.2011&2123.2182&2123.8006&2150.8815&2152.9639&17\\
D$_{\rm J}$(kHz)&0.2099&0.2101&0.1748&0.3679&0.3986&7\\
D$_{\rm JK}$(kHz)&10.0087&9.3279&9.5555&10.1180&-10.8263&9\\
D$_{\rm K}$(kHz)&197.7238&186.2954&187.3222&203.4061&240.6530&3\\
d$_{\rm 1}$(kHz)&-0.0417&-0.0387&-0.0342&-0.0473&-0.0466&4\\
d$_{\rm 2}$(kHz)&0.0053&0.0055&0.0064&0.0054&-0.0005&6\\
\hline
 \hspace{-1.8mm} i-PrCN &&&&\\
A$_{0}$$\rm$(MHz)&7894.7161&7917.9440&7902.1722&7934.5747&7940.8771&31\\
B$_{0}$$\rm$(MHz)&3958.4767&3889.0349&3888.1260&3974.8597&3968.0877&27\\
C$_{0}$$\rm$(MHz)&2886.5734&2861.8154&2861.5579&2899.2361&2901.0532&22\\
D$_{\rm J}$(kHz)&0.5811&0.5890&0.5881&0.5937&0.6102&153\\
D$_{\rm JK}$(kHz)&11.3818&11.1077&11.2111&11.8208&12.1772&42\\
D$_{\rm K}$(kHz)&-4.2564&-4.1421&-4.0748&-4.5264&-5.2324&61\\
d$_{\rm 1}$(kHz)&-0.3437&-0.3346&-0.3443&-0.2403&-0.2440&69\\
d$_{\rm 2}$(kHz)&-0.1652&-0.1760&-0.1792&-0.1860&-0.1892&76\\
\hline
\end{tabular}}}
\footnotesize{The experimental data are taken from \cite{Muller2016} for n-PrCN and \cite{2011Muller} for i-PrCN}
\end{table*}
\begin{figure*}
    \centering
    \includegraphics[width=0.5\textwidth]{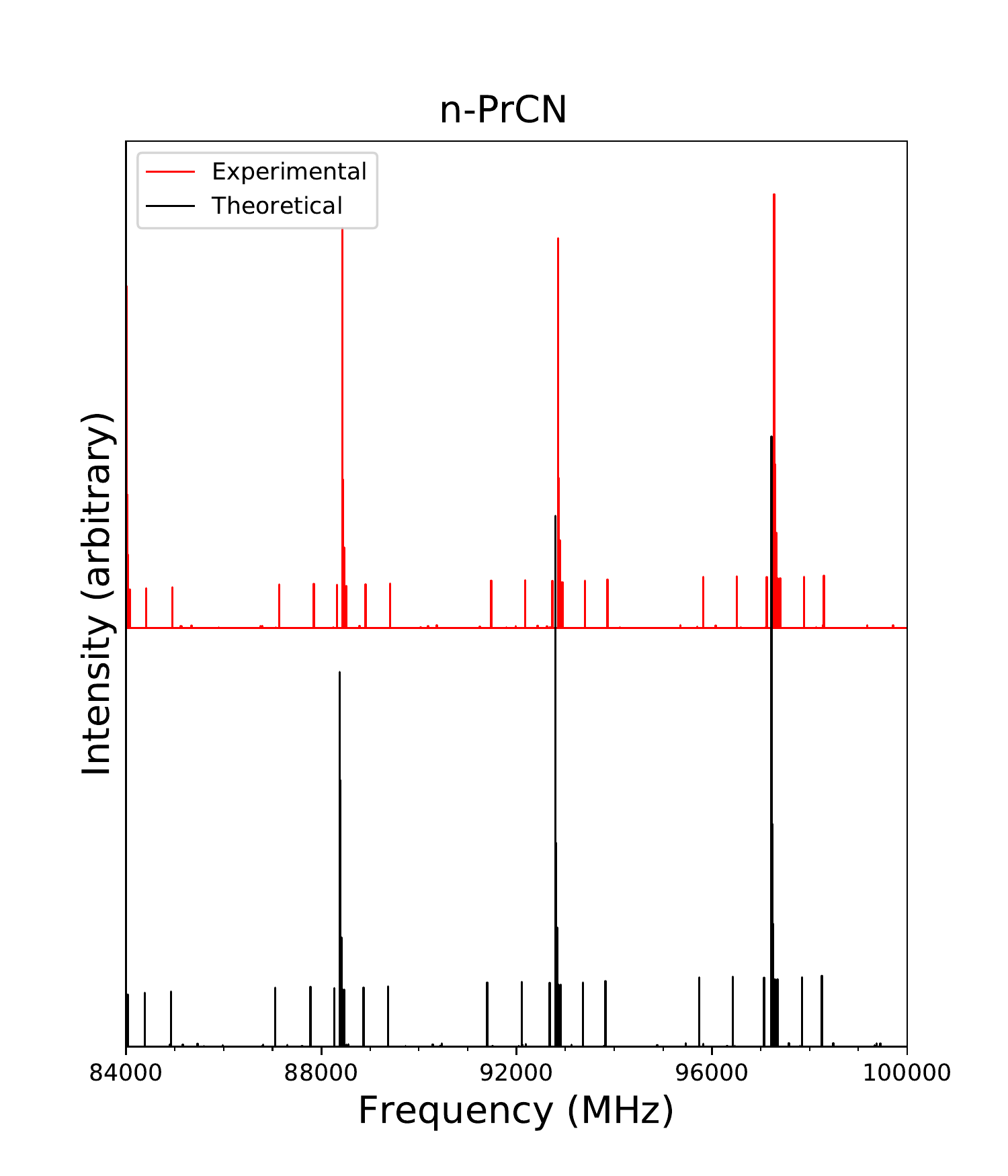}\includegraphics[width=0.5\textwidth]{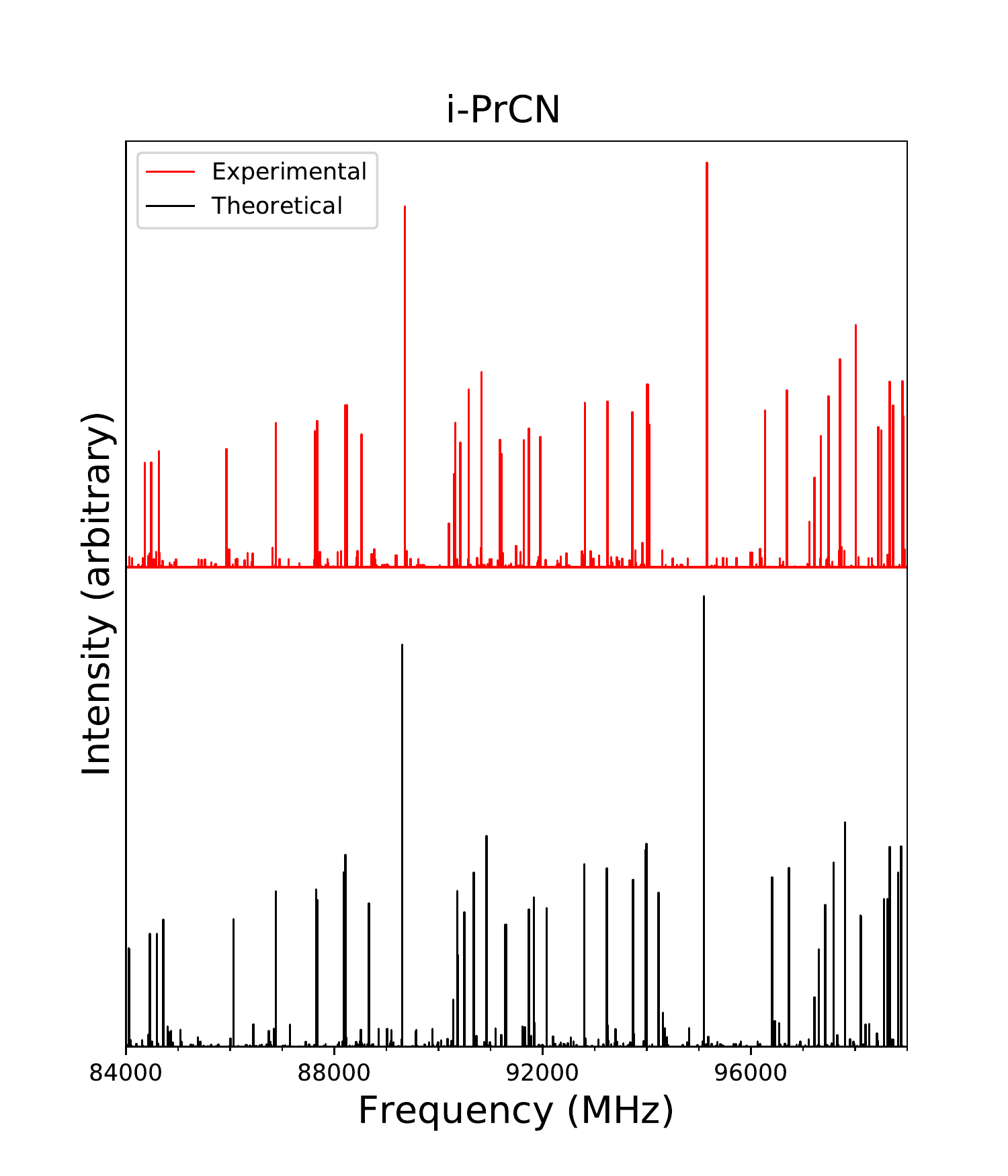}
    \caption{Rotational spectra of n-PrCN and i-PrCN with B3LYP/cc-pVTZ at 300 K (black) in the frequency range of 84-111 GHz. The experimental spectra are shown in red.}
    \label{fig: 5}
\end{figure*}

The results found no conformers for both iso- and neo pentane (Fig. 3 \& 4) due to their higher symmetries. Both iso- and neo-pentane exist in the most stable structures at -60$^\circ$ having comparable energy values of -197.1165 and -197.1220 a.u., respectively. Since iso-pentane is already suggested as the most abundant BCM in Sgr B2 \citep{2017Garrod}, the spectroscopic constants and rotational spectra for iso-pentane are discussed in the following section.

\subsection{\textbf{Efficiency of the computational techniques}}

\subsubsection{Rotational parameters on different basis sets}

The rotational spectroscopic constants of the test molecules were determined using four different levels of theories-- B3LYP/6-311G++(d,p), B3LYP/cc-pVTZ, MP2/cc-pVDZ, and MP2/aug-cc-pVDZ, which are given in Table 1 along with the experimental values for both n-PrCN \citep{Muller2016} and i-PrCN \citep{2011Muller}. The uncertainties in the experiments are also shown in Table 1. \citet{2011Muller} reported the computation of the spectroscopic constants of i-PrCN with B3LYP and MP2 level theories along with the laboratory experiment. The computed rotational constants (A$_{0}$, B$_{0}$, C$_{0}$) and quartic centrifugal distortion constants (D$_{J}$, D$_{JK}$, D$_{K}$, d$_{1}$, d$_{2}$) of both the test molecules \textbf{\textcolor{red}}are showing excellent match with the experiments at the B3LYP level of theory compared to the MP2 (Table 1).

The calculated ground state rotational constants of n-PrCN agree with the experimental values within $\sim$ 0.8 and $\sim$ 1.3 per cent at the B3LYP/aug-cc-pVTZ and B3LYP/6-311G++(d,p), respectively (Table 1). For i-PrCN, the values are within $\sim$ 0.2 per cent and $\sim$ 0.6 per cent respectively. At MP2/cc-pVDZ and MP2/aug-cc-pVDZ, these values are within $\sim$ 2.0 per cent of experiments for both n-PrCN and i-PrCN (Table 1). This concludes that B3LYP/aug-cc-pVTZ is the most suitable level of theory in the determination of rotational constants. 

As can be seen in Table \ref{Tab 1}, the experimental quartic centrifugal distortion parameters agree favorably with the theoretical values at all the levels of computational methods particularly calculated at the B3LYP/aug-cc-pVTZ level. For n-PrCN, the D$_{J}$ value is much closer to the experimental value at B3LYP/aug-cc-pVTZ level with relative error $\sim$ 8 per cent and for i-PrCN $\sim$ 3 per  cent. However, other constants are not showing such similarity with the experimental values. This could be because of the fact that the anti n-PrCN is much closer to the prolate symmetric top limit, which makes the determination of A$_{0}$, D$_{K}$, etc. difficult from a-type transitions alone \cite{Muller2016}). 
In the case of i-PrCN, these constants are much more accurate at B3LYP/aug-cc-pVTZ. 

After the calculation of rotational constants at all levels of theories with the error within $\sim$ 8 per cent of the experimental values. The results calculated at the B3LYP/aug-cc-pVTZ are consistent with the previous studies \citep{2011Muller}. Therefore, the B3LYP/aug-cc-pVTZ level is found the most suitable level for the determination of rotational parameters of BCMs.
\begin{table*}
\centering
    \caption{Rotational transitions of ground state of n-propyl cyanide(n-PrCN) and iso-propyl cyanide(i-PrCN) with relative error ($\sim \delta$)}
    \begin{tabular}{ccccccc}
    \hline
    \hline
    {Transition}&\multicolumn{2}{c}{Frequency}&&Relative error($\delta$)&Experimental Uncertainty \\
    \cline{1-3}
      $J^{'}_{K^{'}_{a},K^{'}_{c}}$ $\to  J_{K_{a},K_{c}}$ &Experimental&Calculated&& \\
      &(MHz)&(MHz)&&($\%$)\\
      \hline
       \hspace{-1.8mm} n-PrCN &&&&\\
       $20_{3,17}$$\to19_{3,16}$&88516.6823&88257.2017&&0.29&1\\
      $21_{3,19}$$\to20_{3,18}$&92915.5930&92646.4909&&0.29&1\\
       $22_{3,20}$$\to21_{3,19}$&97345.3458&97063.2126&&0.28&1\\
       $23_{3,21}$$\to22_{3,20}$&101775.4529&101480.2967&&0.28&1\\
       $24_{3,22}$$\to23_{3,21}$&106205.8551&105897.6919&&0.29&1\\
       $25_{4,21}$$\to24_{4,20}$&110611.8632&110290.6091&&0.29&1\\
       \hline
        \hspace{-1.8mm} i-PrCN &&&&\\
        $15_{0,15}$$\to14_{0,14}$&89353.2582&89306.5219&&0.05&1\\
        $16_{0,16}$$\to15_{0,15}$&95154.9314&95104.5702&&0.05&1\\
        $16_{1,15}$$\to15_{1,14}$&99818.3299&99783.1229&&0.04&1\\
        $17_{0,17}$$\to16_{0,16}$&100956.7195&100902.7298&&0.05&1\\
        $15_{5,11}$$\to14_{5,10}$&105647.4322&105568.9042&&0.07&1\\
        $18_{0,18}$$\to17_{0,17}$&106758.5792&106700.9591&&0.05&1\\
        \hline
       \end{tabular}
    \label{tab:2}
\end{table*}

\subsubsection{Simulation of rotational spectra of n-PrCn and i-PrCN}

Figure \ref{fig: 5} illustrates the simulation of the pure rotational spectra of n-PrCN and i-PrCN at 300 K with B3LYP/aug-cc-pVTZ between 84--111 GHz frequency range. Some of the strongest rotational transitions in the considered frequency range of both the test molecules are summarised in Table \ref{tab:2} together with the relative error between the theoretical and experimental values. \textbf{\textcolor{red}}{The uncertainties observed in the experimental frequencies are also given in Table \ref{tab:2}.}

The calculated transition frequencies with B3LYP/aug-cc-pVTZ for n-PrCN and i-PrCN are accurate to within $\sim$ 0.29 and $\sim$ 0.07 per cent of the experimental values, respectively (Table \ref{tab:2} and Fig \ref{fig: 5}). It is clear from Figure 5 and Table 1, even a minor change in rotational constants significantly impacts the precision of spectral simulations.

Only the strongest transitions within the considered frequency range are noted. The R-branch transitions with k$_{a}$ = 3 were the primary focus of the initial transition lines for n-PrCN. These lines, when estimated using theoretical methods, were reasonably close to the experimental values (within $\delta$ $\sim$0.29). For example, the strongest line corresponding to $25_{4,21}$$\to24_{4,20}$ falls at 110290 MHz, demonstrating a closer match with experimental data (110611 MHz). For iso-propyl cyanide, the R-branch transitions with k$_{a}$= 0 exhibit the strongest line corresponds to $18_{0,18}$$\to17_{0,17}$ at 106700 MHz. This transition shows a relative error of $\delta$ = 0.05 with experimental frequency ($\lambda = $106758 MHz).

Based on the above, the B3LYP/aug-cc-pVTZ level of theory is suitable level for calculating rotational constants and simulating rotational spectra for linear and branched chain molecules. The most common isomer of pentane is therefore explored using the B3LYP/aug-cc-pVTZ level of theory in the following section.

\subsection{\textbf{Rotational spectroscopy of iso-pentane}}
\begin{figure}
    \centering
   \includegraphics[width=0.5 \textwidth]{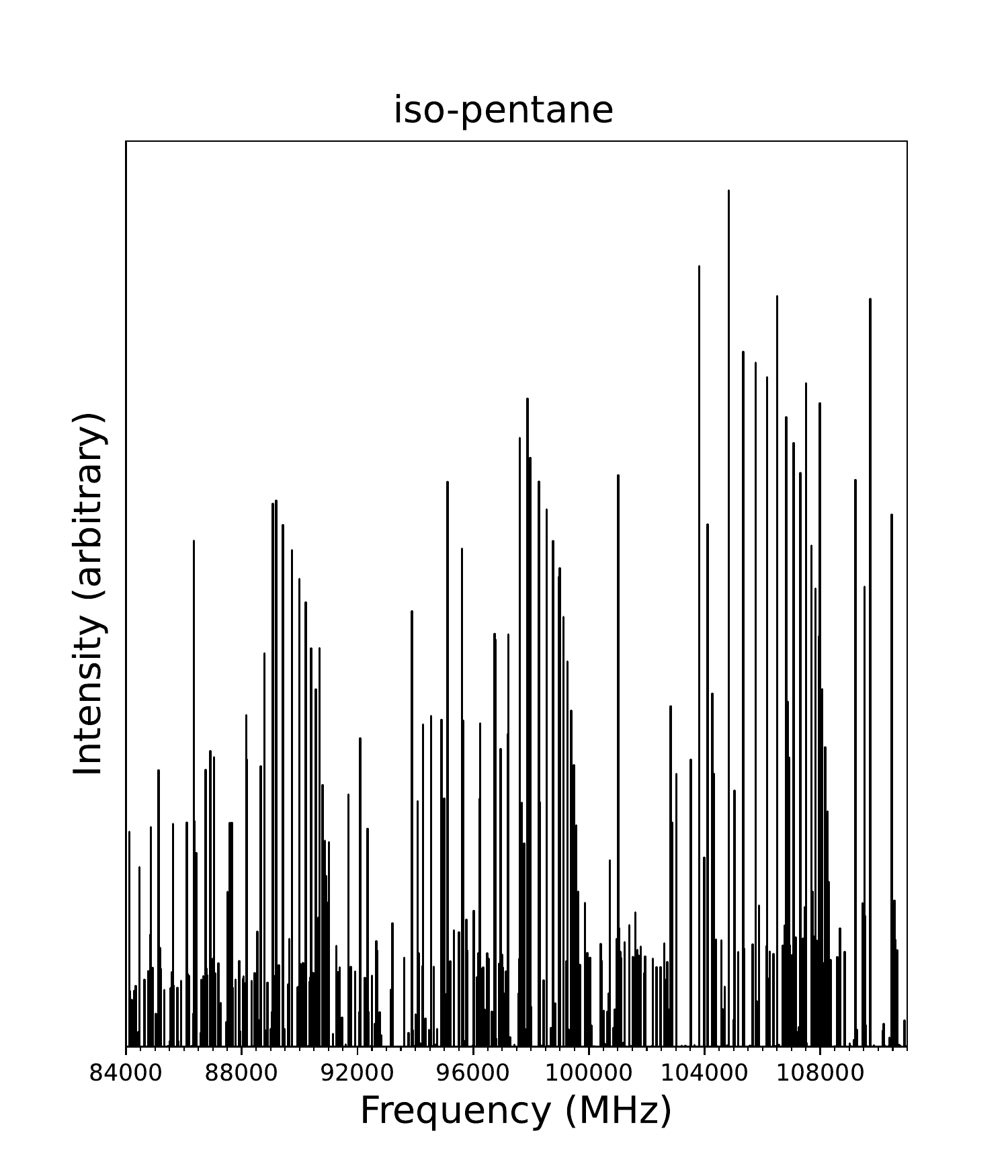}
    \caption{Calculated rotational spectra of iso-pentane at 300 K with B3LYP/cc-pVTZ level}
    \label{fig:6}
\end{figure}

Rotational transitions have so far led to the discovery of the majority of the species in the ISM. In this regard, the molecular constants of iso-pentane have been obtained using the B3LYP/aug-cc-pVTZ level of theory. The pure rotational spectrum of iso-pentane in the frequency range of 84 to 111 GHz at 300 K is simulated (Fig: \ref{fig:6}) using PGOPHER programme utilizing the calculated ground state rotational and quartic centrifugal distortion constants.
\begin{table}
\centering
  \caption{Spectroscopic parameters in the ground state of iso-pentane at B3LYP/aug-cc-pVTZ level}
    \label{tab:3}
    {\begin{tabular}{ccccc}
    \hline
 Parameters&i-C$_{5}$H$_{12}$\\ 
 \hline
A$_{0}$$\rm$ (MHz)&7300.7423\\
B$_{0}$$\rm$ (MHz)&3340.0674\\
C$_{0}$$\rm$ (MHz)&2554.2757\\
D$_{\rm J}$(kHz)&0.5044\\
D$_{\rm JK}$(kHz)&5.5317\\
D$_{\rm K}$(kHz)&-1.0799\\
d$_{\rm 1}$(kHz)&-0.1579\\
d$_{\rm 2}$(kHz)&-0.0597\\
\hline
\end{tabular}}
\end{table}

The rotational constants (A$_{0}$ = 7300, B$_{0}$ = 3340, and C$_{0}$ = 2554) of iso-pentane are small compared to n-PrCN while being similar to i-PrCN. There are 2120 R-branch transitions in total (707 different frequencies as a result of unresolved asymmetric splitting). In its ground vibrational state (v = 0), iso-pentane is assigned with values of J$^{"}$ up to 30 and K$^{"}_{a}$ up to 11 in the millimeter-wave spectrum (84 to 111 GHz).
Some strongest transitions and their corresponding frequencies of iso-pentane are $17_{14,4}$$\to16_{3,13}$; 104691, $8_{7,2}$$\to7_{6,2}$; 103802, $25_{13,12}$$\to16_{13,13}$; 106506 and $9_{7,3}$$\to8_{6,2}$; 109724.
The spectroscopic constants of iso-pentane are tabulated in Table \ref{tab:3}.

The majority of the millimetre range strongest transitions occur at higher frequencies, indicating that these transitions may be helpful in the future for the detection of iso-pentane in high temperature (300 K) region.

\section{{\textbf{Astronomical implications and Conclusions}}}
More than 80 per cent of molecules in the ISM have been identified using pure rotational spectroscopy \citep{2022ApJS..259...30M,2018Agundez,2021Agundez}. Among these, the first detection of a BCM (i-PrCN) had been possible in the last decade. Later on, the modelling results of such astronomical environments found higher abundance of neo-butane and iso-pentane (a few $\times$ 10$^{-10}$n[H$_{2}$]) among all the forms of butane and pentane at all the considered temperatures (20--400 K) \citep{2017Garrod}. Since the current study focuses on the detection probability of BCMs and there is very limited rotational data available for this, we have studied the pure rotational spectra of iso-pentane at 10, 100 and 300 K temperatures. The B3LYP/aug-cc-pVTZ level of theory is most suitable for studying the rotational spectra of BCMs.  
According to the model \citep{2017Garrod}, most complex organic molecules are created on dust-grain surfaces through the formation and addition of smaller radicals, with the end result being that when temperatures rise, the products desorb into the gas phase. The chemical model takes into account mechanisms leading to the formation of C$_{4}$H$_{10}$ on grain surfaces as well as in cold-cloud gas-phase chemistry of straight-chain C$_{4}$ molecules. As a result, during the collapse phase of the model, there is a modest accumulation of n-C$_{4}$H$_{10}$ on the grains. Pentane starts to develop on the dust grains as the temperature of the warm-up phase of the model approaches 20 K; the necessary radicals for pentane formation are produced by cosmic ray-induced photodissociation of either type of C$_{4}$H$_{10}$ with a subsequent addition of a methyl group. The model further explains that the hydrogenation of C$_{4}$H$_{8}$ favours the development of the secondary-radical form of C$_{4}$H$_{9}$ and makes it possible for the iso form to be created in a somewhat higher amount compared to the regular form. By adding a methyl group to the main or secondary form of the C$_{4}$H$_{9}$ radical, both normal and iso-pentane are created on the grain surfaces.

\begin{table}
    \centering
    \caption{Strongest rotational transitions of iso-pentane at various temperature}
    \begin{tabular}{cccccc}
    \hline
    \hline
    {Transition}&{Frequency}&& &T\\
      $J^{'}_{K^{'}_{a},K^{'}_{c}}$ $\to  J^{"}_{K^{"}_{a},K^{"}_{c}}$ &&& \\
      &(MHz)&&&(K)\\
      \hline
      $7_{7,1}$$\to6_{6,1}$&97874.2899&&&10\\
      $17_{14,4}$$\to16_{13,4}$&217869.2493&&&100\\
      $21_{21,0}$$\to20_{20,0}$&302295.3474&&&300\\

      \hline
       \hline
       \end{tabular}
    \label{tab:4}
\end{table}

\begin{figure}
    \includegraphics[width=0.50\textwidth]{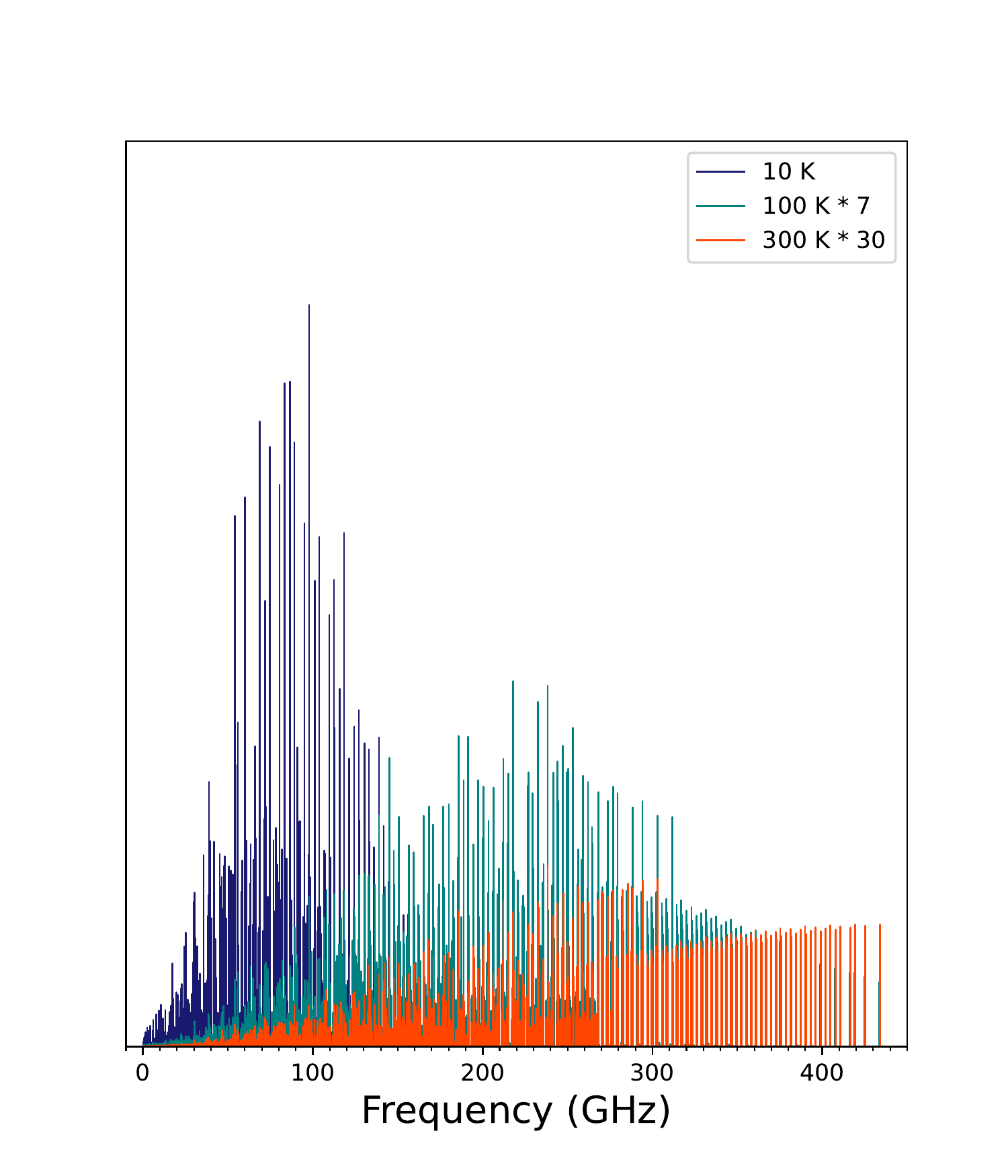}
    \caption{Calculated rotational spectrum of iso-pentane at 10 K
(purple), 100 K (teal), and 300 K (orange). For the purposes of display, these relative intensities for the 100 K plot have increased by a factor of 7, while for the 300 K plot, the increase is 30.}
    \label{fig:7}
\end{figure}

Despite the fact that iso-pentane forms on grain surfaces and is abundant and stable at all temperatures, the astronomical detection of iso-pentane has not been possible in the most prominent molecular clouds in space as yet. There might be two distinct reasons for this. First, a number of higher BCMs are not characterised at all and, second, they are at a resolution that is too low to enable an astronomical search. In light of this, we are considering the simulation of iso-pentane at three different temperatures 10 K, 100 K and 300 K (Fig \ref{fig:7}).
The strongest transitions are tabulated at all the temperatures considered in Table \ref{tab:4}. The perusal of Table 4 and Figure 6 show that the strongest feature at 10 K corresponds to J$^{"}$ = 6 at 98 GHz. At 100 K, the strongest line fall near 218 GHz (J$^{"}$ = 16). The strongest transition falls in the range of 200–250 GHz when temperature starts to increase from 100 to 300 K. The strength of individual transitions is dramatically reduced when temperature rises, which might be due to the greater partition function in this range (Figure 7).

Since the probable detection frequency range of the BCMs falls within 84--111 GHz, the astronomical search of iso-pentane seems most feasible in low temperature regions corresponding to J$^{"}$ = 6 at 98 GHz. With extremely accurate measurements of the rest frequencies and associated spectroscopic constants, the iso-pentane search in the ISM may now be conducted with a high level of assurance.

\section*{Acknoledgment}

AP (Anshika Pandey) acknowledges Banaras Hindu University and UGC, New Delhi, India, for providing a fellowship. SS acknowledges fellowship from Banaras Hindu University. AV acknowledges research fellowship from DST SERB (SERB-EMR/2016/005266) and UGC. AP (Amit Pathak) acknowledges financial support from the IoE grant of Banaras Hindu University (R/Dev/D/IoE/Incentive/2021-22/32439), financial support through the Core Research Grant of SERB, New Delhi (CRG/2021/000907) and thanks the Inter-University Centre for Astronomy and Astrophysics, Pune for associateship. KAPS acknowledges the UGC Faculty Recharge Program of Ministry of Human Resource Development (MHRD), Govt. of India and University Grants Commission (UGC), New Delhi and support from IoE grant of Banaras Hindu University.

\bibliography{cas-refs}





\end{document}